\documentclass{jfm}
\usepackage{graphicx}
\usepackage{psfrag}
\usepackage{natbib}

\usepackage{amsmath}
\usepackage{amssymb}

\newcommand\cD{{\cal D}}

\title[Floating of Multiple Interfacial Objects]{Equilibrium Conditions for the Floating of Multiple Interfacial Objects}
\author[D. Vella, P. D. Metcalfe \& R. J. Whittaker]
{D\ls O\ls M\ls I\ls N\ls I\ls C\ns V\ls E\ls L\ls L\ls A%
,\ns
P\ls A\ls U\ls L\ns D.\ns M\ls E\ls T\ls C\ls A\ls L\ls F\ls E\break
\and R\ls O\ls B\ls E\ls R\ls T\ns J.\ns W\ls H\ls I\ls T\ls T\ls A\ls K\ls E\ls R}

\affiliation{Institute of Theoretical Geophysics, Department of Applied Mathematics and Theoretical Physics, University of Cambridge, Wilberforce Road, Cambridge, CB3 0WA, U.\ K.}

\begin{document}

\maketitle

\begin{abstract}
  We study the effect of interactions between objects floating at
  fluid interfaces, for the case in which the objects are primarily
  supported by surface tension.  We give conditions on the density and
  size of these objects for equilibrium to be possible and show that
  two objects that float when well-separated may sink as the
  separation between the objects is decreased.  Finally, we examine
  the equilbrium of a raft of strips floating at an interface, and
  find that rafts of sufficiently low density may have infinite
  spatial extent, but that above a critical raft density, all rafts
  sink if they are sufficiently large.  We compare our numerical and
  asymptotic results with some simple table-top experiments, and find
  good quantitative agreement.
\end{abstract}

\section{Introduction}

A common table-top demonstration of the effects of surface tension is
to float a metal needle horizontally on water: even though the density
of the needle is much greater than that of water, the needle is able
to remain afloat because of the relatively large vertical component of
surface tension. This effect is a matter of life or death for
water-walking insects \cite[][]{Bush}, and is also important in
practical settings such as the self-assembly of small metallic
components into macroscopic structures via capillary flotation forces
\cite[][]{Whitesides}.  In this engineering setting an object
should not only float when isolated at the interface, but
must also remain afloat after it has come into contact with other
interfacial objects, and portions of the meniscus that supported it
have been eliminated.  Although the interactions that cause
interfacial objects to come into contact and form clusters have been
studied extensively \cite*[see, for
example,][]{Mansfield,Kralchevsky,Vella2}, the implications of such
interactions on the objects' ability to remain afloat have not been
considered previously.

Here we consider the effects of these interactions via a series of
model calculations that shed light on the physical and mathematical
concepts that are at work in such situations.  For simplicity, the
calculations presented here are purely two-dimensional, though the
same physical ideas apply to three-dimensional problems.

\section{Two horizontal cylinders}

\begin{figure}
\centering
\includegraphics[height=3.5cm]{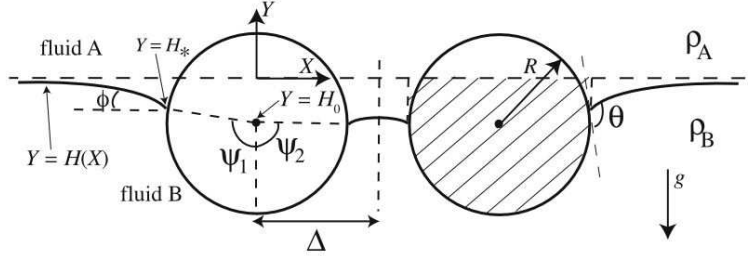}
\caption{Cross--section of two parallel, horizontal cylinders lying at an interface with a non-dimensional centre--centre separation of $2\Delta$.}
\label{2cylchan}
\end{figure}

Perhaps the most natural way to characterise the effects of
interaction is to ask how the maximum vertical load that can be
supported by two floating cylinders varies as the distance between
them is altered. We thus consider two cylinders of infinite length
lying horizontally at the interface between two fluids of densities
$\rho_A<\rho_B$, as shown in figure \ref{2cylchan}. We assume that
these cylinders are non-wetting so that the contact angle $\theta$, a
property of the three phases that meet at the contact line, satisfies
$\theta>\pi/2$.

We non-dimensionalise forces per unit length by the surface tension
coefficient, $\gamma_{AB}$, and lengths by the \textit{capillary
  length}, $\ell_c\equiv(\gamma_{AB}/(\rho_B-\rho_A)g)^{1/2}$, and use
non-dimensional variables henceforth. We wish to determine the maximum
weight per unit length, $W$, that can be supported by each of two
identical cylinders with radius $R$ and centre--centre separation
$2\Delta$.

To remain afloat each individual cylinder must satisfy a condition of
vertical force balance: their weight (or other load) must be balanced
by the vertical contributions of surface tension and the hydrostatic
pressure acting on the wetted surface of the cylinder. We assume that an external horizontal force is applied to maintain the separation of the cylinders and so do not consider the balance of horizontal forces explicitly.

Using the notation of figure \ref{2cylchan}, the vertical force balance
condition may be written $W=U_1+U_2$ where
\begin{equation}
U_i\equiv-\sin(\theta+\psi_i)-H_0R\sin\psi_i+
\tfrac{1}{2} R^2 (\psi_i+\sin\psi_i\cos\psi_i)\hspace{0.5cm} (i=1,2),
\label{vfbal}
\end{equation}
are the contributions to the vertical upthrust provided by the
deformation on each half of the cylinder separately, and $H_0$ is the
height of the cylinders' centres \emph{above} the undeformed free
surface. Physically, the first term on the right hand side of
\eqref{vfbal} is the vertical component of surface tension, and the
second and third terms quantify the resultant of hydrostatic pressure
acting on the wetted perimeter of the cylinder. The latter is given by
the weight of water that would fill the dashed area in figure
\ref{2cylchan} \cite[see][]{Keller}.

The angles $\psi_1$ and $\psi_2$ are determined by the interfacial
shape, which is governed by the balance between hydrostatic pressure
and the pressure jump across the interface associated with interfacial
tension. This balance is expressed mathematically by the
Laplace--Young equation. In two dimensions this is
\begin{equation}
H_{XX}=H (1+H_X^2)^{3/2},
\label{LapYoung}
\end{equation}
where $H(X)$ is the deflection of the interface (again measured positive
upwards) from the horizontal, and subscripts denote differentiation.
Since the exterior meniscus
extends to infinity, the first integral of \eqref{LapYoung} is particularly simple in this instance and allows the height of the contact line, $H_*$, to be related to the interfacial inclination, $\phi$, via
\begin{equation}
\cos\phi=1-\tfrac{1}{2}H_*^2.
\label{geomconst}
\end{equation}
This, together with the geometrical condition $\phi=\theta+\psi_1-\pi$, allows $\psi_1$ to be eliminated from \eqref{vfbal} in favour of
$H_0(=H_*+R\cos\psi_1)$ and $\theta$.

For the interior meniscus, we simultaneously obtain
$\psi_2$ and the shape $H(X)$, by using the  MATLAB routine \texttt{bvp4c} to solve the nonlinear eigenproblem
\begin{equation}
  \label{eq:1}
  \begin{aligned}
    H_{XX} &= H (1+H_X^2)^{3/2},\\
    H_{X}(R \sin \psi_2) &= \tan(\theta+\psi_2),\\
    H(R\sin\psi_2) &= H_0 - R \cos \psi_2, \\
    H_X(\Delta) &= 0,\\
  \end{aligned}
\end{equation} on $[R\sin\psi_2,\Delta]$.

\begin{figure}
\centering
\includegraphics[height=4cm]{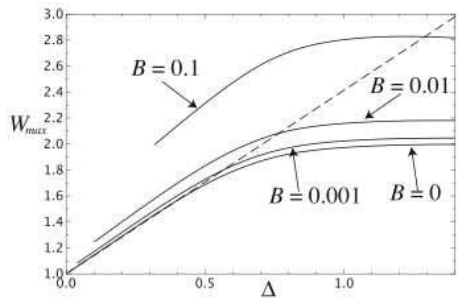}
\caption{Numerical results for the maximum load that can be supported
  by a single cylinder in the presence of another a distance $2\Delta$
  away when $\theta=2\pi/3$ for several values of the Bond number, $B\equiv R^2$. The dashed line shows the linear approximation \eqref{2cylasy} for
  the limiting case $B=0$ when $\Delta\ll1$.}
\label{modgraph}
\end{figure}

With the angles $\psi_1$ and $\psi_2$ calculated, $W(H_0)$ can be
determined from \eqref{vfbal}, and the maximum load that can be
supported, $W_{\mathrm{max}}$, can be found numerically by varying
$H_0$.  Of
particular interest is the dependence of $W_{\mathrm{max}}$ on the
cylinder separation, which is shown for several values of the
\emph{Bond number} $B\equiv R^2$ in figure \ref{modgraph}. This plot
includes the limiting case $B=0$, corresponding to the application
of two point forces to the interface.

The results presented in figure \ref{modgraph} show that as the
distance between two cylinders decreases, the maximum vertical load
that can be supported by each cylinder decreases.  Physically, this
result is intuitive since even though the interior meniscus is not
completely eliminated in this instance, the vertical force that this
meniscus can exert on the cylinder is diminished by the symmetry
requirement that $H_X(\Delta)=0$. In particular, for small $B$ and
$\Delta$ the total force that can be supported by each cylinder is around half of that which can be supported by an isolated cylinder. This corresponds to the simple
physical picture that for small Bond number, the restoring force is
supplied primarily by the deformation of the meniscus \cite[][]{Hu};
when the interior meniscus is eliminated, the contact line length per cylinder, and
hence the force that surface tension can provide, are halved. From
this we expect that very dense objects that float when isolated at an
interface might sink as they approach one another. Since floating
objects move towards one another due to capillary flotation forces
\cite[see][for example]{Mansfield}, it seems likely that this effect
may be ubiquitous for dense objects floating at an interface and may
also have practical implications.

For $B=0$ we can compute the asymptotic form of $W_{\mathrm{max}}$ for
$\Delta\ll1$ by noting that for small separations the interior
meniscus has small gradients and the Laplace--Young equation
\eqref{LapYoung} may be approximated by $H_{XX}=H$, which has the
solution $H(X)=H_0\cosh(X-\Delta)/\cosh\Delta$.  Thus, the vertical
force provided by the deformation is
$W=-H_0(\tanh\Delta+(1-{H_0}^2/4)^{1/2})$, which is extremised when
$({H_0}^2-2)/(4-{H_0}^2)^{1/2}=\tanh\Delta$. Choosing the real root of
this quartic corresponding to a maximum in $W$ and making consistent
use of $\Delta\ll1$, $W_{\mathrm{max}}$ can be expanded as a series in
$\Delta$. We obtain
\begin{equation}
W_{\mathrm{max}}=1+\sqrt{2}\Delta+O(\Delta^2),
\label{2cylasy}
\end{equation}which compares favourably with the numerically computed results
presented in figure \ref{modgraph}.

\section{Two touching strips\label{2strip}}

\begin{figure}
\centering
\includegraphics[height=2cm]{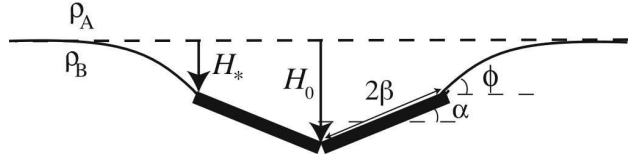}
\caption{Cross--section of two shallow, touching strips floating at a liquid--fluid interface.}
\label{2psetup}
\end{figure}

Whilst the scenario considered in the previous section may be relevant
in practical situations, it does not lend itself to particularly
simple experimental validation. To allow for such a comparison, we now consider the
equilibrium of two infinitely long, shallow strips of dimensional
thickness $\ell_ct$, width $2\ell_c\beta$, and density $\rho_s$, floating with
their long edges in contact so that the interior meniscus is
completely eliminated. The configuration is shown schematically in figure \ref{2psetup}.
Here, we are no longer bound by a contact angle condition but instead
assume that the meniscus is pinned to the uppermost corners of the
strips.  The additional complication of the strip's angle of
inclination to the horizontal, $\alpha$, is determined by the balance
of torques. (This condition is satisfied automatically for shapes
with circular cross-section and constant contact angle, as shown by
\cite{Singh}.)

\begin{figure}
\centering
\psfrag{labelbeta}{$\beta$}
\psfrag{labeld}{$\cD$}
\includegraphics[height=5cm]{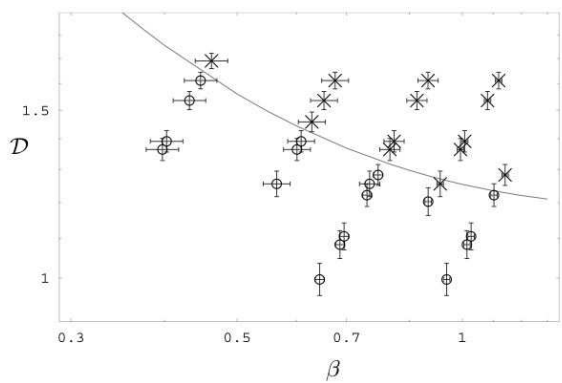}
\caption{Numerically computed values of $\cD_{\mathrm{max}}$ as the half-width of the strips, $\beta$, is varied (solid line). Experimental results (as described in text) are shown by points $\times $ (strips that sink)  and $\bigcirc$ (strips that float).}
\label{2ptouch}
\end{figure}

Equating moments about the point of contact (thereby eliminating
the need to calculate the tension force that the strips exert on one
another) and balancing vertical forces, we obtain the conditions for
equilibrium
\begin{eqnarray}
\cD\beta\cos\alpha=\sin(\phi-\alpha)-\beta(H_0+\tfrac{4}{3}\beta\sin\alpha)\label{2ptorbal},\\
\cD\beta=\tfrac{1}{2} \sin\phi -\beta\cos\alpha(H_0+\beta\sin\alpha),
\label{2pvfbal}
\end{eqnarray}
where 
\begin{equation}
\cD\equiv \frac{(\rho_s-\rho_B)t}{\rho_B-\rho_A}
\label{Ddefn}
\end{equation}
is the appropriate ratio of the density of the strips to those of the surrounding
fluids. After eliminating $\cD$ between \eqref{2ptorbal} and \eqref{2pvfbal} and using \eqref{geomconst} with the relation $H_*=H_0+2\beta\sin\alpha$ to eliminate $\phi$, we have a single equation for $\alpha$ given
particular values of $\beta$ and $H_0$. Thus, for fixed $\beta$ and a given value
of $H_0$, we may solve for $\alpha$ and deduce the corresponding value
of $\cD$ from \eqref{2pvfbal}. By varying $H_0$ we are then able to
calculate the maximum value of $\cD$ for which equilibrium is
possible, much as before. The numerical results of this calculation
are presented in figure \ref{2ptouch}.

Also shown in figure \ref{2ptouch} are experimental results showing
points in $(\beta,\cD)$ parameter space for which two identical strips
remained afloat or sank upon touching. These experiments were
performed with strips of stainless-steel shim of length
$69\,\mathrm{mm}$ with $\rho_s=7905\,\mathrm{kg}\,\mathrm{m}^{-3}$ and
thickness $0.4$ or $0.5\,\mathrm{mm}$. These were floated on aqueous
solutions of $0\%$, $10\%$ or $25\%$ methanol in air (so that $\rho_A
\ll \rho_B$), allowing a wide range of values of $\beta$ and $\cD$ to
be probed. The strips were then allowed to come into contact naturally
via the mutually attractive flotation force \cite[][]{Mansfield}. The
data are plotted with horizontal and vertical error bars. The former
indicate the uncertainty in the measurement of the strip widths. The
latter indicate the uncertainty in the additional vertical force
contribution of the ends (since the strips are of finite length),
which may be shown to be equivalent to an uncertainty in the effective
value of $\cD$. The agreement between our experiments and theory in
this instance is very good.

\section{The floating of a flexible raft}

By adding additional strips to a floating pair of strips, a flexible
raft is formed. Given the analysis of the preceding sections it is
natural to expect that as the raft is lengthened in this manner, there
will come a point where its weight (which scales with its total
length) exceeds the force that can be supplied by surface tension
(which is constant) and so the raft should sink. The situation is
complicated by the fact that the raft may bow in its middle,
displacing a considerable amount of liquid in this region, as pointed
out by \cite{Saif}. We now address the question of whether, for
a raft of given weight per unit length, there is a maximum raft length
before sinking occurs.

\begin{figure}
\centering
\psfrag{labelbeta}{$\beta$}
\psfrag{labeld}{$\cD$}
\includegraphics[height=2.5cm]{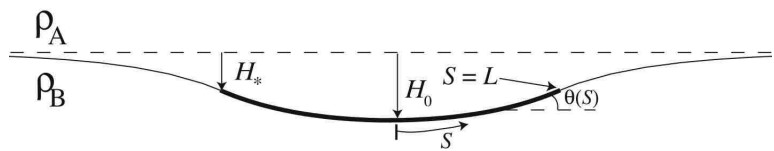}
\caption{Schematic illustration of a flexible raft floating at an interface.}
\label{raftsetup}
\end{figure}

We tackle this problem by treating the raft as a continuum, shown schematically in fig.~\ref{raftsetup}, and
formulating an equation for the deformation of such a raft. This
generalises the linear analysis of \cite{Mansfield} and allows us to
consider situations in which interfacial deformations are no longer
small, including the existence of a threshold length for sinking.

\subsection{Governing equation}

We use a variational approach to determine the shape $(X(S),H(S))$ of
the raft and the surrounding meniscus, though the same result may also
be obtained by considering the force balance on an infinitesimal raft
element. The non-dimensional arc-length, $S$, is measured from the
raft's axis of symmetry at $S=0$, with the two ends of the raft being
at $S=\pm L$.  For simplicity, we neglect the intrinsic bending stiffness of the raft, although \cite{Vella} have shown that interfacial rafts do, in general, have some resistance to bending. The variational principle states that raft
shapes must minimise the energy of the system over variations in
$H(S)$ and $X(S)$, subject to the constraint that $X_S^2+H_S^2=1$.
Introducing a Lagrange multiplier $\lambda(S)$ associated with this
constraint, we find that equilibrium raft shapes extremise
\begin{equation}\label{eq:energy}
{\cal E}\equiv \int_{-\infty}^\infty\left(X_S(H^2/2-1)+
  \cD H\chi+(1-\chi)+\lambda(S)\bigl[(X_S^2+H_S^2)^{1/2}-1\bigr]\right)\mathrm{d}S,
\end{equation}
where $\cD$ was defined in \eqref{Ddefn} and
\begin{equation}
  \chi(S)\equiv
  \begin{cases}
    1, & |S| \le L \\
    0, & |S| > L,
  \end{cases}
\end{equation} 
is the indicator function of the raft.

The first two terms in the integral \eqref{eq:energy} correspond to
the gravitational energy of the displaced fluid and the raft, the
third term is the surface energy of the uncovered liquid area, and the
final term ensures that the constraint $X_S^2+H_S^2=1$ is satisfied.
Note that a small increase in arc-length such that $X_S^2+H_S^2>1$
increases the energy of the system so that the Lagrange multiplier
$\lambda(S)$ may be interpreted physically as the tension in the
raft/meniscus. That the raft can support a tension at all may seem
counterintuitive. It is a consequence of the attractive
capillary interaction that would exist between two infinitesimally separated raft elements.

Requiring ${\cal E}$ to be stationary with respect to variations in
$H(S)$ and $X(S)$ yields differential equations for $X$ and $H$. Using
the differential form of the constraint, $X_SX_{SS}+H_SH_{SS}=0$, we
may eliminate $\lambda$ to obtain $\lambda_S=\chi \cD H_S$. This may
be integrated using the boundary term from integration by parts at
$\pm\infty$, the boundary conditions $H(\pm\infty)=0$ and
$X_S(\pm\infty)=1$ as well as the continuity of $\lambda$ at the raft
edge, $S=\pm L$, to give $\lambda=1+\chi\cD(H-H_*)$, where $H_*\equiv H(\pm
L)$.
We now find the raft shape numerically by solving the nonlinear
eigenproblem
\begin{equation}
  \begin{gathered}
    X_S=\cos\theta,\quad
    H_S=\sin\theta,\quad
    \theta_S=\frac{H+\cD\cos\theta}{1+\cD(H-H_*)},\\
    X(0) =0,\quad 
    \theta(0) = 0, \quad
    \theta(L) = 2\arcsin(H_*/2),\quad
    H(L) = H_*,
  \end{gathered}
\label{goveq}
\end{equation}
for $X(S)$, $H(S)$, $\theta(S)$ on $[0,L]$, and $H_*$, using the MATLAB routine
\texttt{bvp4c}. The results of this computation may be verified by calculation of the quantity
\begin{equation}
P(\theta)\equiv\tfrac{1}{2}H^2+\bigl[1+\cD(H-H_*)\bigr]\cos\theta-1,
\label{cons_q}
\end{equation} which is conserved and, from the boundary conditions, equal to $0$.

In the limit of small deformations \eqref{goveq} reduces to the
simpler linear form studied by \cite{Mansfield} in the context of
determining typical raft profiles. Here, however, we wish to determine
whether a maximum raft length, $2L_{\mathrm{max}}$, exists and if so
find its value for a raft of given density $\cD$. To investigate this,
small deformation theory is inadequate since sinking is an essentially
non-linear phenomenon.

The symmetry condition $\theta(0)=0$ ensures that $H_*\geq-\cD/2$ and
that $H_0\equiv H(0)\geq-\cD$, so that the centre of the raft may sink
at most to its neutral buoyancy level. In what follows, it will be
convenient to treat $H_0$ and $\cD$ as parameters giving rise to a
particular raft semi-length $L(H_0,\cD)$; we find
\begin{equation}
L(H_0,\cD)=\frac{H_0^2}{2\cD}\int_0^1\frac{2+H_0^2(y-1)}{\Bigl\{\bigl[2+ H_0^2(y-1)\bigr]^2-\bigl[2-(H_0+y H_0^2/2\cD)^2\bigr]^2\Bigr\}^{1/2}}\mathrm{d}y,
\label{intlenexp}
\end{equation}
which follows by changing integration variables from $S$ to $H$ in
$L=\int_0^L\mathrm{d}S$. This allows us to consider the behaviour of
$L$ for a given value of $\cD$ as $H_0$ is varied.

The tension at the midpoint of the raft is given by
$1-H_0^2/2$, showing that the raft goes into
compression if $H_0\leq-\sqrt{2}$. Physically this is unrealistic,
corresponding to a divergence in $\theta_S$. If $\cD<\sqrt{2}$, this
situation is avoided automatically since $H_0\geq-\cD>-\sqrt{2}$ but
for $\cD\geq\sqrt{2}$ we must consider this possibility; we therefore
consider these two cases separately.

\subsection{The case $\cD<\sqrt{2}$}

\begin{figure}
\centering
\includegraphics[height=4cm]{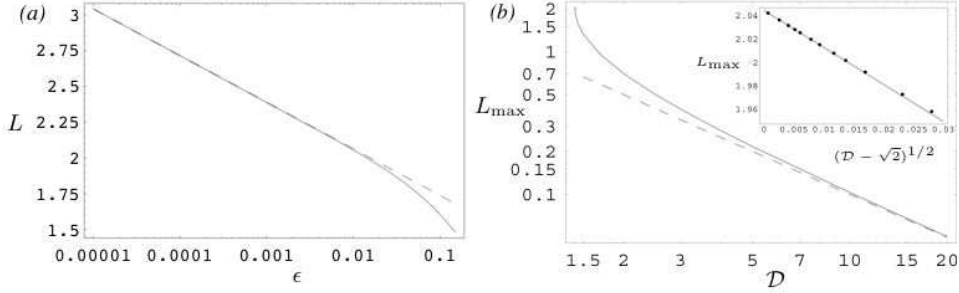}
\caption{($a$) Numerical results of the calculation of $L$ as a function of
  $\epsilon\equiv H_0+\cD$ (solid line) compared to the asymptotic result
  \eqref{smalld} for $\epsilon\ll1$ (dashed line) for the case $\cD=1.4$. ($b$) Main figure: Numerical results of the calculation of
  $L_{\mathrm{max}}$ as a function of the density ratio $\cD\geq
  \sqrt{2}$ (solid line), together with the large ${\cD}$ asymptotic result $L_{\mathrm{max}}\sim 1/\cD$ (dashed line). Inset:
  Rescaled graph comparing the numerically computed values (points) of
  $L_{\mathrm{max}}$ with the asymptotic expansion \eqref{sqrtasy}
  (solid line).}
\label{DRes}
\end{figure}

When $\cD<\sqrt{2}$, the centre of the raft may reach its neutral
buoyancy depth $H_0=-\cD$ without going into compression. Numerical
computation of the integral \eqref{intlenexp} suggests that rafts grow
arbitrarily long as $H_0\searrow-\cD$ (see figure \ref{DRes}$a$). To
show that this is the case, we consider the asymptotic behaviour of
the integral \eqref{intlenexp} in the limit $\epsilon\equiv\cD+H_0\ll1$\footnote{Note
  that $\epsilon\geq0$, since $H_0\geq-\cD$.}. This is done by
splitting the range of integration into two sub-regions $[0,\delta]$
and $(\delta,1]$, where $\delta$ is unspecified save for the condition
that $\epsilon\ll\delta\ll1$ \cite[see][]{Hinch}. Within these two
regions, the two integrands may be simplified using approximations
compatible with this gearing of $\delta$, and the resulting integrals 
evaluated analytically. Upon expanding these results for $\delta\ll1$
the leading order terms in $\delta$ cancel, yielding
\begin{eqnarray}
L=-\mu\log\left(\frac{\epsilon}{\cD}\right)&+&\mu\log\left(\frac{8\mu^2}{\sqrt{2}\mu(7+\mu^2)^{1/2}+4-\cD^2}\right)\nonumber\\
&-&2\cD\arctan\left(\frac{\cD(3\mu-\sqrt{2}(7+\mu^2)^{1/2})}{3\cD^2+\sqrt{2}\mu(7+\mu^2)^{1/2}}\right)+O(\epsilon^{1/2}),
\label{smalld}
\end{eqnarray}
where $\mu\equiv(1-\cD^2/2)^{1/2}$. This result compares favourably with the numerical results in figure \ref{DRes}(a). In particular, notice that $L$ diverges logarithmically as $H_0\searrow -\cD$ (i.e.\ as
$\epsilon\rightarrow0$) so that rafts of arbitrary length are
possible. It also interesting to note that \eqref{smalld} may be inverted to give an estimate of
$H_0=-\cD+\epsilon$ for given values of $\cD$ and $L$ --- a useful
result when calculating raft shapes for large $L$.

That a raft of sufficiently low density can grow arbitrarily large in
horizontal extent without sinking seems surprising at first glance.
However, as new material is added to the raft, it may be accommodated at its neutral buoyancy level 
without the raft going into compression.  Therefore, the raft's ability to remain afloat is not jeopardised and it is almost obvious that these low density rafts may grow arbitrarily long without sinking. 

\subsection{The case $\cD\geq\sqrt{2}$}

In this case, the raft cannot reach its neutral buoyancy level,
invalidating the argument just given to explain why, with
$\cD<\sqrt{2}$, rafts may be arbitrarily large. We thus expect that a
maximum raft length does exist and, further, that the limiting raft
has $H_0=-\sqrt{2}$.  Numerical computation of $L$ as a function
of $H_0$ indicates that a critical half-length $L_{\mathrm{max}}$ does
exist, but that it is not attained with exactly this value of $H_0$. Instead,
there is a competition between the raft sinking deep into the liquid
(to support its weight by increased hydrostatic pressure) and having
its ends a large distance apart (i.e. lower pressure but over larger
horizontal distances), and some compromise is reached. Given the abrupt change in behaviour observed as $\cD$ increases past $\sqrt{2}$, we are particularly interested in the nature of this transition. Numerical computations suggest that for $\eta^2\equiv\cD-\sqrt{2}\ll1$,
$L_{\mathrm{max}}$ occurs when $H_0=-\sqrt{2}+c\eta^2$ for some
constant $c$.  Motivated by this observation, we let
$H_0=-\sqrt{2}+c\eta^2$ and again split the domain of integration in
\eqref{intlenexp} into two regions $[0,\delta']$ and $[\delta',1]$
where $\eta^2\ll\delta'\ll1$. This allows us to calculate $L$
to leading order in $\eta$, yielding
\begin{equation}
L= 2\sqrt{2}\arctan\left(\frac{\sqrt{7}}{3}\right)+\eta \frac{2^{3/4}c}{(c+1)^{1/2}}\left[\mathrm{K}\left(\frac{c+2}{2(c+1)}\right)
-\frac{2(c+1)}{c}\mathrm{E}\left(\frac{c+2}{2(c+1)}\right)\right]+O(\eta^2),
\label{Lexp}
\end{equation}
where
$\mathrm{K}(k)\equiv\int_0^{\pi/2}(1-k^2\sin^2\phi)^{-1/2}\mathrm{d}\phi$
and
$\mathrm{E}(k)\equiv\int_0^{\pi/2}(1-k^2\sin^2\phi)^{1/2}\mathrm{d}\phi$
are the complete elliptic integrals of the first and second kinds,
respectively. The coefficient of $\eta$ in \eqref{Lexp} has a maximum
for fixed $\eta$ at $c=c^*$, where $c^*$ satisfies
\begin{equation}
\mathrm{K}\left(\frac{c^*+2}{2c^*+2}\right)=2\mathrm{E}\left(\frac{c^*+2}{2c^*+2}\right).
\end{equation}
Hence $c^*\approx0.5332$, and we obtain the asymptotic expression
\begin{equation}
L_{\mathrm{max}}= 2\sqrt{2}\arctan\left(\tfrac{\sqrt{7}}{3}\right)-3.1525\left(\cD-\sqrt{2}\right)^{1/2}+O\left(\cD-\sqrt{2}\right),
\label{sqrtasy}
\end{equation}
which compares very favourably with the numerically computed values of
$L_{\mathrm{max}}$ presented in the inset of figure \ref{DRes}($b$).

For the limiting case $\cD=\sqrt{2}$, the above analysis breaks down
since then $\eta=0$ and we lose the freedom to vary $H_0$. However, by
letting $\epsilon=c\eta^2$ (so that $H_0=\epsilon-\sqrt{2}$) we take
the limit $\eta\rightarrow0$ of \eqref{Lexp} with $\epsilon\ll1$ fixed
to find
\begin{equation}
L(\epsilon)= 2\sqrt{2}\arctan\left(\tfrac{\sqrt{7}}{3}\right)+\epsilon^{1/2}2^{3/4}\left[\mathrm{K}(\tfrac{1}{2})
-2\mathrm{E}(\tfrac{1}{2})\right]+O(\epsilon).
\label{etazero}
\end{equation}
This has a maximum value of $2\sqrt{2}\arctan(\sqrt{7}/3)$ at
$\epsilon=0$, which is the same value as that found from
\eqref{sqrtasy} in the limit $\cD\searrow \sqrt{2}$. It is also
reassuring to note that, as $\cD\nearrow\sqrt{2}$ with $\epsilon$
fixed, the expression in \eqref{smalld} also gives $L=
2\sqrt{2}\arctan(\sqrt{7}/3)+ O(\epsilon^{1/2})$.

For completeness, we consider finally the limit $\cD\gg1$. To leading order in $\cD^{-1}$, the integral for $L(H_0,\cD)$ is given by
\begin{equation}
L(H_0,\cD)\sim
\cD^{-1} \int_{1-H_0^2/2}^1\frac{u}{(u^2-(1-H_0^2/2)^2)^{1/2}}\mathrm{d}u=
\cD^{-1} H_0(1-H_0^2/4)^{1/2}.
\end{equation} This has a maximum value of $\cD^{-1}$ at $H_0=-\sqrt{2}$ so that in
the limit $\cD\gg1$, $L_{\max}\sim\cD^{-1}$. This is precisely as we
should expect physically since large density objects can only float
when the contribution of surface tension dominates that of the
buoyancy due to excluded volume and, in particular, it must balance
the weight of the raft. This asymptotic result compares favourably with the the numerical results presented in figure \ref{DRes}($b$).

\subsection{Comparison with experiment}

\begin{figure}
\centering
\includegraphics[height=4cm]{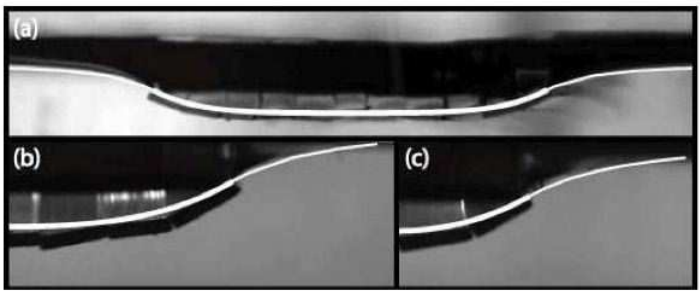}
\caption{Comparison between experimental realisation of a
  two-dimensional raft (viewed through the side of a transparent tank)
  and the theoretically predicted shape (superimposed white line). The
  rafts float at an air--water interface and have varying values of
  $\cD$ and $L$: ($a$) a complete raft with $\cD=1.02$,
  $L=4.03$ ($b$) one half of a raft with $\cD=1.27$, $L=1.47$ and
  ($c$) one half of a raft with $\cD=1.27$, $L=2.57$. The typical
  width of each individual strip is $2$ mm. The black region apparently
  above the raft is in fact a reflection of the black base of the
  confining tank from the meniscus at the edge of the tank}
\label{raft}
\end{figure}

A direct comparison between the theoretical results outlined so far
and experimental results is difficult since we have modelled the raft
as a perfectly flexible continuum body of infinite extent along its
axis of symmetry. Despite these limitations, the theoretical raft
shapes calculated via this model are in good agreement with simple
experiments in which thin strips of stainless steel shim are laid
side-by-side at an air--water interface, as shown in figure \ref{raft}
--- even when the raft consists of only a small number of strips and
we might not expect the continuum approximation to be valid.

Although this agreement is encouraging, our main interest lies more in
whether there is a maximum length for such a raft to remain afloat, as
predicted by the model. Practical considerations mean it is difficult
to produce strips of stainless steel shim narrower than about $2$~mm
in the workshop, so the comparisons we are able to draw between our
model and experiments can only be semi-quantitative.  In spite of
these limitations, we find that for stainless steel strips of length
$69$ mm and thickness $0.5$ mm the maximum raft-length is $4-6$ mm for
an aqueous solution of 25\% methanol (so that $1.645\geq\cD\geq1.580$)
and $6-8$ mm for 15\% methanol (so that $1.494\geq\cD\geq1.424$),
which are certainly consistent with the corresponding theoretical
results of $4.6$ mm $\leq L_{\mathrm{max}}\leq4.8$ mm and $6.5$ mm
$\leq L_{\mathrm{max}}\leq7.2$ mm, respectively. Here the length was
increased by floating additional strips near the raft and allowing
them to come into contact via the mutually attractive capillary
flotation forces until the raft was no longer stable and sank. With
$\cD=1.02$ and $\cD=1.27$, we were able to add many strips without any
sign of the raft sinking indicating that this process might be
continued indefinitely.

\section{Discussion}

In this article, we have quantified the conditions under which objects
can remain trapped at a fluid-fluid interface, and shown that when the
deformation of the meniscus is suppressed by the presence of other
objects the supporting force that can be generated decreases
dramatically. For two small, parallel cylinders or strips, the maximum
force that can be supported close to contact is only that provided by
the contribution from the exterior meniscus and so sufficiently dense
objects sink upon contact. A two-dimensional raft of touching,
floating strips may compensate partially for this loss of meniscus by
sinking lower into the fluid. For $\cD<\sqrt{2}$, this effect allows
rafts of arbitrary length to remain afloat. For $\cD\geq\sqrt{2}$,
there is a maximum length (dependent on $\cD$) above which equilibrium is not possible.

Although the agreement between the experiments and theory presented here
is good, our analysis was confined to two dimensions, whereas
experiments must be carried out in the three-dimensional world.
Similarly, we have limited ourselves to considering the
\emph{equilibrium} of objects at an interface. We are currently
studying the dynamics of sinking for the case of two touching strips considered in
section \ref{2strip},  and find that a simple hydrodynamic model produces good agreement with experiments.

\begin{acknowledgements}
  We are grateful to David Page-Croft for his help in the laboratory and Herbert Huppert for comments on an earlier draft.  DV and RJW are supported by the EPSRC. PDM gratefully acknowledges
  the financial support of Emmanuel College, Cambridge.
\end{acknowledgements}

 \end{document}